\documentstyle[preprint,pra,aps]{revtex}
\begin{document}
\draft
\title{\bf
Induced Parity Nonconserving Interaction
and Enhancement of Two-Nucleon Parity\\
Nonconserving Forces \footnote{Phys.Rev. {\bf C}, to appear}
}
\author{V.~V.~Flambaum and O.~K.~Vorov}
\address{School of Physics, University of New South Wales,
Sydney, 2052,NSW, Australia}
\date{September 14}
\maketitle
\begin{abstract}
Two-nucleon parity nonconserving (PNC) interaction induced by the
single-particle PNC weak potential and the two-nucleon residual strong
interaction is considered.
An approximate analytical formula for this Induced PNC Interaction (IPNCI)
between proton and neutron is
derived ($Q({\bf r} {\bf \sigma}_{p} \times
{\bf \sigma}_{n})
\delta({\bf r}_{p}-{\bf r}_{n})$),
and the interaction constant is estimated.
As a result of coherent contributions from the nucleons to the PNC
potential,
IPNCI is an order of magnitude stronger
($\sim A^{1/3}$)
than the residual
weak two-nucleon interaction and has a different coordinate and
isotopic structure (e.g., the strongest part of IPNCI does not contribute
to the PNC mean field).
IPNCI plays an important role in the formation of PNC effects, e.g.,
in neutron-nucleus reactions.
In that case, it
is a technical way to take into account the
contribution of the distant
(small) components of a compound state which dominates the result.
The absence of such
enhancement ($\sim A^{1/3}$)
in the case of T- and P-odd interaction completes the picture.
\end{abstract}

\section{Introduction}
\label{sec:level1}

The parity nonconserving (PNC) nucleon interaction in nuclei
and PNC effects in neutron-nucleus reactions
are subject of current interest for both experimentalists and theorists
\cite{te0}-\cite{MITCHEL}.
The values of the PNC effects depend on the weak interaction matrix elements
between
compound states.
Usually two sources of the PNC effects are discussed: a
single-particle weak potential
$w$ which describes the interaction of a nucleon with a weak mean
field of the nucleus and a residual
two-particle weak interaction $W$. In principle, the
matrix elements of $w$ and $W$ should be
calculated with respect to the
eigenstates of the strong interaction
Hamiltonian. However, in practice some truncated basis set of states is used
to describe physical states at excitation energies less
than the gap between single-particle shells.
For example, in the description of nuclear compound states
and the P-odd admixtures in them \cite{te1},\cite{we}
it is natural
to include into the basis set only ``principal''
components which have energies close
to the energy of the compound state and dominate the normalization sum.

The number of such components is already about $10^{6}$
in a compound state. However, it is still not enough
since these components consist from the valence (incomplete)
shell orbitals only (see e.g., Ref. \cite{BM}) and do not
contain opposite parity orbitals with the same angular
momentum (these orbitals belong to different shells).
Thus, matrix element of the single particle weak potential
$w$ between compound states is zero in the `` principal component''
approximation since it can mix these opposite parity orbitals
only \cite{Zar},\cite{kad}. To avoid this problem one should
consider admixture of the distant small components which
contain the necessary opposite parity orbitals from other
shells. Any transfer of a particle from the valence shell
to another one gives rise to an excitation energy
$E_{sp} \sim 5...8$ MeV which is much more than a typical
matrix element of the residual strong interaction $V_{S}$.
Therefore, one can easily admix a small component to compound
states using perturbation theory in $V_{S}$:
\begin{equation}
|c> \quad = \quad |c_{0}> \quad + \quad \sum_{\alpha}
\frac{<\alpha |V_{S}|c_{0}>}{E_{c} - E_{\alpha}} |\alpha>.
\end{equation}
where $|c_{0}>$ is the ``principal part'' of the compound state.
Now we can calculate the matrix element of $w$ between
the opposite parity compound states $|s>$ and $|p>$:
\begin{equation}
<s|w|p> \quad = \qquad
\sum_{\alpha}
\frac{<s_{0}|w|\alpha><\alpha|V_{S}|p_{0}> +
<s_{0}|V_{S}|\alpha><\alpha|w|p_{0}>}
{E_{0}-E_{\alpha}}.
\end{equation}
The single-particle weak potential $w$ can be written down
in the form
\begin{eqnarray}
\hat w(1) =
\frac{Gg^{W}}{2\sqrt{2}m}\lbrace({\bf \sigma}{\bf p})\rho +
\rho ({\bf \sigma}{\bf p})\rbrace
\simeq \frac{\xi}{m} {\bf \sigma p},
\nonumber\\
 \qquad \qquad \varepsilon=1.0 \cdot 10^{-8}g^{W},
\xi = \frac{Gg^{W}}{2\sqrt{2}} \rho_{0} ,
\qquad \rho_{0} = \frac{2 p_{F}^{3}}{3 \pi^{2}}
\end{eqnarray}
where $G=10^{-5}m^{-2}$ is the Fermi constant, $m$ is the
nucleon mass, ${\bf p}$ and ${\sigma}$ are the nucleon momentum
and doubled spin,
$\rho$ is the nuclear density, $\rho \simeq \rho_{0}= const$
inside the nucleus. We can use relations
\begin{equation}
p=i m [H,r], \qquad \qquad H |\alpha> = E_{\alpha} |\alpha>,
\end{equation}
to calculate sum over $\alpha$ in Eq.(2).
Here $H$ is a Hamiltonian of the system (here we neglect
the spin-dependent part of $H$). Using closure relation
$\sum_{\alpha} |\alpha><\alpha| = 1$ we obtain
\begin{equation}
<s|w|p> \quad =
\quad i <s_{0}| [\sum_{k} \xi_{k} \sigma_{k}
{\bf r}_{k},V_{S}]|p_{0}>,
\end{equation}
the sum is taken over nucleons [actually only the nucleons near the
Fermi surface contribute to this sum (see Ref. \cite{we} and below)].

If we introduce the effective interaction (the Induced Parity
Nonconserving Interaction, IPNCI)
\begin{equation}
V^{IPNCI}= i [\sum_{k} \xi_{k} \sigma_{k}
{\bf r}_{k},V_{S}]
\end{equation}
we need not refer to the small components and calculate
the matrix elements of the IPNCI between the ``principal''
components of the compound states only (see Eq.(5)).

To derive formula (5) for IPNCI we used some approximations
(constant nuclear density, and spin-independent Hamiltonian $H$).
When doing numerical calculations these approximations
are not necessary. In our work \cite{we} we have used a more
accurate perturbation theory expression for the matrix
elements of the IPNCI between the nuclear orbitals $a,b,c,d$:
\begin{equation}
V^{IPNCI}_{abcd}=
\sum_{i} \left[ \frac{w_{ai}V_{S,ibcd}}{\epsilon_{a}-\epsilon_{i}}
 - \frac {V_{S,aicd}w_{ib}}{\epsilon_{i}-\epsilon_{b}} +
\frac{w_{ci}V_{S,abid}}{\epsilon_{c}-\epsilon_{i}} -
\frac{V_{S,abci}w_{id}}{\epsilon_{i}-\epsilon_{d}} \right].
\end{equation}
Using the approximation $\rho = \rho_{0}$ in Eqs.(3),(4), the
Eq.(7) can be reduced to Eq.(6) for the operator of the IPNCI.

The approximate analytical expression (6) is convenient to study
coordinate, spin, and isospin structure and also the strength of the
IPNCI. It will be shown that the IPNCI is an order of magnitude
stronger than the residual two-particle weak interaction $W$.
This amplification ($\sim A^{1/3}$) can be explained by
a coherent contribution of all the nucleons to the PNC potential
which induces the IPNCI. As a result the IPNCI gives the dominating
contribution to the matrix elements of the weak interaction
between the compound states and determines the value of the PNC
effects in nucleus-neutron reactions.

The natural question arises: we obtained the enhancement in
the treatment of the residual strong interaction to first order
of perturbation theory.
Will
this enhancement ``survive'' in ``all-order'' treatment?
To answer this question we will present in the next
chapter the derivation of the IPNCI which is not based on the
perturbation theory treatment of the residual strong
interaction.

\section{Derivation of IPNCI. Unitary transformation}
\label{sec:level2}

We start with the nuclear Hamiltonian $H$ in the form
\begin{equation}
H=H_{0} \quad + \quad V_{S} \quad + \quad W \quad + \quad F,
\end{equation}
where the first term $H_{0}={\bf p}^{2}/2m+U_{S}(r)$ is the single particle
Hamiltonian of the nucleons with inclusion of the single-particle part of
the strong interaction $U_{S}(r)$ (strong potential), $V_{S}$ is the residual
two-body strong interaction,
$F$ describes other possible interactions, e.g., coupling to
electromagnetic field, anapole moment operator \cite{A1}-\cite{A3} {\it etc}.
The operator
$W=\hat W(1,2)$ is the two-body weak PNC interaction,
\cite{Nog}-\cite{nov3}:
\begin{eqnarray}
\hat W (1,2) \quad = \quad \frac{G}{\sqrt{2}} \frac{1}{2m}
((g^{W}_{12}{\bf \sigma}_{1}- g^{W}_{21}{\bf \sigma}_{2})
\nonumber\\
\times \lbrace
({\bf p}_{1}-{\bf p}_{2}) \delta({\bf r}_{1}-{\bf r}_{2})+
\delta({\bf r}_{1}-{\bf r}_{2})({\bf p}_{1}-{\bf p}_{2}) \rbrace+
                                     \nonumber\\
g'^{W}_{12} [{\bf \sigma}_{1} \times {\bf \sigma}_{2}] \nabla_{1}
\delta({\bf r}_{1}-{\bf r}_{2}) ) \qquad \qquad \qquad \qquad \qquad
\end{eqnarray}
where $G=10^{-5}m^{-2}$ is the Fermi constant, $m$ is the nucleon mass,
${\bf p}$ and ${\sigma}$ are the nucleon momentum and its doubled spin
respectively
\footnote{This weak Hamiltonian goes back to works by Feynman and Gell-Mann
\cite{GELLMANN}; the constants $g$ in it were the subject of numerous studies
(see e.g. \cite{MCKELLAR},\cite{ADELBERGER}, and references therein).
We used the values of these constants from Refs. \cite{DDH},\cite{nov3}.}
(hereafter, the notation ${\bf a} \times {\bf b}$
means exterior vector product).
The nucleon
dimensionless constants $g_{p,n}$
(see e.g. Refs. \cite{Nog}-\cite{nov22})
are of the order of unity and may be chosen in such
a way that only direct terms in (9) should be accounted for.

It is well known
(see e.g. \cite{nov3},\cite{STODOLSKY})
that the main P-odd effects caused by the
weak interaction $\hat W$ in (1) are usually due to
the effective one-body P-odd
interaction, or the ``weak potential'', $\hat w(1)$,
acting on the nucleon $1$,
which arises from
averaging $\hat W(1,2)$ over the states  of the nucleon $2$
(see Eq.(3) for $w \equiv w(1) = \langle W(1,2) \rangle$).
The weak potential
constants $g^{W}_{p},g^{W}_{n}$
are given by
$g^{W}_{p}  =\frac{Z}{A} g^{W}_{pp}  + \frac{N}{A} g^{W}_{pn},
g^{W}_{n} =  \frac{Z}{A} g^{W}_{np} + \frac{N}{A} g^{W}_{nn} $
for proton and neutron respectively.
(Now, the  notation $\varepsilon \simeq 1.0 \cdot 10^{-8}g^{W}$
is widely used).
The coherent contribution from all the paired
nucleons
yields the nuclear density $\rho$ in the expression (3).

As it has been mentioned above
the coherent single-particle P-odd contribution (3) does not work
effectively in mixing of
the nearest excited nuclear states.
Therefore, the P-odd effects in this energy region can be determined
by the purely two-particle ``residue'', $:\hat W(1,2):$ of
the weak interaction
$\hat W(1,2)$, given by the
difference
\begin{equation}
:\hat W(1,2): \quad \equiv
\hat W(1,2) -  \langle \hat W(1,2) \rangle
= \hat W(1,2) -  \hat w(1),
\end{equation}
which does not contain coherent summation in contrast to (3).

As mentioned above, the purpose of this work is to show that
the residual
strong interaction $V_{S}$ in the Hamiltonian (8) gives rise to
appearance of an {\it effective P-odd two-particle interaction}
(IPNCI)
which turns out to be stronger than the initial one, $:\hat W(1,2):$.
We show that IPNCI contains the enhancement of order $\sim A^{1/3}$ times,
compared to the initial two-particle P-odd term, and, moreover, the additional
enhancement can arise \cite{PRC}, if the residual parity conserving strong
interaction
contains momentum-dependent structures \cite{Brown},\cite{SAPERSHTEIN},
\cite{BACKMAN}.
The latter is taken into account by the solving the equation  for the
effective field, what is equivalent to summation of the infinite sum of
graphs, analogous to that considered in the Theory
of Finite Fermi System (TFFS) \cite{Migdal},\cite{Brown}.

We start with the case when the strong interaction $V_{S}$ is ``switched off''.
As is known from Refs.\cite{nov1},\cite{nov2},
in the simple model of a constant nuclear density $\rho \simeq \rho_{0}=
2p_{F}^{3}/3\pi^{2}$
it is easy to find the result of the action of the perturbation
$\hat w(1)$
\begin{eqnarray}
\tilde \psi =
exp(-{\hat a}) \psi^{0}
\simeq (1-i\xi {\bf \sigma r})\psi^{0}, \qquad {\hat a}=i\xi {\bf \sigma r}
                              \nonumber\\
\xi=\frac{G}{\sqrt{2}}g^{W}\rho_{0}=\varepsilon m,
\qquad \xi = \xi_{0}+\xi_{\tau} \tau_{z}
\end{eqnarray}
where $\psi^{0}$ is the unperturbed wave function, and $\tau_{z}=-1(+1)$ is
isospin projection for proton(neutron). To get this solution, one
should also neglect spin-orbit interactions.
Accordingly, the matrix elements of any operator $O$, including the
Hamiltonian, can be calculated by using the unperturbed wave functions
$\psi^{0}$ and the transformed operator $\tilde O$:
\begin{eqnarray*}
\langle \tilde \psi_{a}|O|\tilde \psi_{b} \rangle=
\langle  \psi_{a}^{0}|\tilde O| \psi_{b}^{0} \rangle=
\langle  \psi_{a}^{0}|e^{\hat a} O e^{-\hat a}| \psi_{b}^{0} \rangle
\nonumber\\
\simeq \langle  \psi_{a}^{0}|O + [{\hat a},O]| \psi_{b}^{0} \rangle,
\end{eqnarray*}
where $e^{\hat a} \equiv e^{i \xi ({\bf \sigma r})}$ is the operator
of the corresponding unitary transformation with the single-particle
anti-Hermitian $\hat a$. Correct choice of the transformation
yields
compensation of the single-particle P-odd potential in the Hamiltonian
$e^{\hat a} H e^{-\hat a}$: ${\hat w} + [{\hat a},H_{0}] = 0$.
The effect of this potential is now included into the renormalized operators
${\tilde O}$ rather than the wave functions ${\tilde \psi}$.

Let us switch on the strong interaction $V_{S}$ and
seek now for an operator $e^{{\hat A}}$
with the renormalization
resulting from $V_{S}$ taken into account.
(Eventually, as we will see below the operator ${\hat A}$
differs from ${\hat a}$ mainly
due to the renormalization
of the weak interaction constant by the residual strong interaction $V_{S}$.)
The transformed Hamiltonian looks like:
\begin{eqnarray}
\tilde H \quad = \quad e^{\hat A}H e^{-{\hat A}} \quad =
\qquad \qquad \qquad \qquad \qquad
\nonumber\\
H_{0} \quad + \quad V_{S} \quad + \quad \hat w \quad +
\quad :\hat W:  \quad
+\quad [{\hat A},H_{0}] \quad
\nonumber\\
+ \quad F \quad +
\quad  [{\hat A},F] \quad + \quad [{\hat A},V_{S}]
\end{eqnarray}
where we have used the decomposition (10) and neglected all terms above
the first order in the weak interaction.
To obtain the effective two-particle P-odd interaction acting
in the valence shells we should find  the
operator ${\hat A}$ in such a way that the single-particle P-odd contribution
in $e^{\hat A}He^{-{\hat A}}$ will be compensated. The last term in (12) is
a two-body operator. We employ the same decomposition, as in (10):
$[{\hat A},V_{S}] \equiv \langle [{\hat A},V_{S}] \rangle +
:[{\hat A},V_{S}]:$, where the
first single-particle term is the average over the paired nucleons,
and the second one, $:[{\hat A},V_{S}]:$,
which yields zero under such averaging,
is the effective induced two-particle interaction which we are seeking
for:
\begin{equation}
V_{IPNCI}= \quad :[{\hat A},V_{S}]: \quad , \qquad \langle V_{IPNCI} \rangle
\equiv 0.
\end{equation}
Thus, if we require the ``compensation equation''
\begin{equation}
\hat w  \quad + \quad [{\hat A},H_{0}] \quad +
\quad \langle [{\hat A},V_{S}] \rangle =0
\end{equation}
to be fulfilled, the transformed Hamiltonian takes the form
\begin{equation}
\tilde H = H_{0} \quad + \quad V_{S} \quad + \quad :\hat W: \quad +
\quad V_{IPNCI} \quad
+ \quad F \quad + \quad [{\hat A},F],
\end{equation}
where no single-particle P-odd potential is present.
Thus, there are three sources of the parity nonconservation in Eq.(15):

1) the commutator $[{\hat A},F]$ which gives a direct contribution
of the PNC potential $w(1)$
to the matrix elements of an external field $F$
( $\langle \psi | F + [{\hat A},F] | \psi ' \rangle =
\langle {\tilde \psi} | F | {\tilde \psi '} \rangle$);

2) The residual two-body weak interaction $:W:$ ;

3) $V_{IPNCI}$, which plays the same role as $:W:$, but is
enhanced in comparison with $:W:$ (see below).

To solve the equation (14) and find the explicit form of the IPNCI we use
the Landau-Migdal \cite{Brown},\cite{Migdal},\cite{Landau},\cite{SAPERSHTEIN}
parametrization of the strong
interaction:
\begin{equation}
V({\bf r}_{1},
{\bf r}_{2})=C\delta({\bf r}_{1}-{\bf r}_{2})[f+f'{\bf \tau}_{1}{\bf \tau}_{2}+
h{\bf \sigma}_{1}{\bf \sigma}_{2}+h'{\bf \tau}_{1}{\bf \tau}_{2}
{\bf \sigma}_{1}{\bf \sigma}_{2}],
\end{equation}
where $C=\frac{\pi^{2}}{p_{F}m}=300$ $MeV \times fm^{3}$ is the universal
Migdal constant \cite{Migdal},\cite{Brown},\cite{SAPERSHTEIN} and the
strengths $f,f',h,h'$ are in fact functions of $r$ via density dependence:
$f=f_{in}-(f_{ex}-f_{in})(\rho(r)-\rho(0))/\rho(0)$ (the same for $f',h,h'$).
(Quantities subscripted by ``in'' and ``ex''
characterize interaction strengths in the depth of the nucleus and on
its surface, respectively). This interaction goes backwards to Landau Fermi
liquid theory \cite{Landau}. With its parameter values
listed below, it has been successfully used by many authors
(see Refs. \cite{Brown}) to quantitatively describe many
properties of heavy nuclei.
The conventional choice widely used for heavy nuclei is (see \cite{Migdal},
\cite{Brown},\cite{SAPERSHTEIN}):
$f_{ex}=-1.95$, $f_{in}=-0.075$ $f'_{ex}=0.05$ $f'_{in}=0.675$,
$h_{in}=h_{ex}=0.575$, and $h'_{in}=h'_{ex}=0.725$.
It is easy to check that, in the same approximation of constant density
as used above, the operator ${\hat A}$ is proportional to ${\hat a}$:
${\hat A} = i {\tilde \xi} ({\bf \sigma r})$.
Evaluating the commutator in (13,14), we obtain
\begin{eqnarray}
[{\hat A},V_{S}] =
-2 {\tilde \xi}_{\tau}C\delta({\bf r}_{1}-{\bf r}_{2}) \biggl\{ (h'-h)
(\tau_{2z}-\tau_{1z}) {\bf r}_{1} \cdot {\bf \sigma}_{2}
\times {\bf \sigma}_{1}
+
(h'-f')(\tau_{2} \times \tau_{1})_{z}({\bf \sigma}_{2} - {\bf \sigma}_{1},
{\bf r}_{1})
\biggl\}
\qquad \qquad \qquad
\nonumber\\
\langle [{\hat A},V_{S}] \rangle = 0; \quad V^{IPNCI}(1,2)= [{\hat A},V_{S}].
\qquad \qquad \qquad \qquad \qquad \qquad
\end{eqnarray}
Since the last term in the compensation equation (14)
is zero in this case, the operator ${\hat A}$ coincides with ${\hat a}$
and the values of the
constants ${\tilde \xi}_{\tau}$ just coincide with their ``bare'' values
$\xi_{\tau}$ (11) (i.e.,
without the strong interaction).
The first term in Eq.(17) induces transitions $pn \rightarrow pn$,
while the second one $pn(np) \rightarrow np(pn)$.
For contact interactions,
the second term (which is in fact an exchange term in comparison with the first
one)
can be reduced to the first term using Fierz transformation (see e.g.
\cite{OKUN}). After this transformation,
\begin{equation}
\hat V^{IPNCI}(1,2) = \frac{1}{2} Q \delta({\bf r}_{1}-{\bf r}_{2})
({\bf \tau}_{2z}-
{\bf \tau}_{1z}){\bf r}_{1}{\bf \sigma}_{2} \times {\bf \sigma}_{1}
\rightarrow \qquad
Q {\bf r}_{p}({\bf \sigma}_{p} \times {\bf \sigma}_{n}) \delta({\bf r}_{p}-
{\bf r}_{n}),
\end{equation}
\begin{displaymath}
Q= -2(\xi_{p}-\xi_{n})(f'-h)C = \frac{4}{3}\frac{p_{F}^{2}}{m}
\frac{G}{\sqrt{2}}(g^{W}_{p}-g^{W}_{n})(f'-h).
\end{displaymath}
We stress that this expression is valid within the nucleus only
(recall that $Q \sim \xi \sim \rho$).
When using this expression one has to assume the exchange term $pn \rightarrow
np$ is excluded. However, the conventional choice of Landau-Migdal
interaction constants already corresponds to the same assumption.
This means
that the second term in the expression (17) for the IPNCI should be simply
omitted (to avoid double counting) and the final expression for the IPNCI
includes
$pn-pn$ interaction only, i.e. the constant of the IPNCI is
\begin{equation}
Q= -2(\xi_{p}-\xi_{n})(h'-h)C=
\frac{4}{3}\frac{p_{F}^{2}}{m}
\frac{G}{\sqrt{2}}(g^{W}_{p}-g^{W}_{n})h_{pn},
\end{equation}
$h_{pn}=h-h'$ is the constant of the residual strong proton-neutron
spin-flip interaction.
This problem with the definition of the IPNCI constant is due to the fact
that the Landau-Migdal interaction is a phenomenological effective interaction
rather than the
{\it ab initio} strong interaction. For example, it can contain ``fictitious''
spin dependence coming from the Fierz transformation of the exchange term with
the spin-independent interaction $C \delta({\bf r}_{1}-{\bf r}_{2})$. However,
this ``fictitious'' spin dependence does not contribute to
the IPNCI since in the case
of an initial spin-independent interaction the Fierz transformation gives
$h'-h = f'-h = 0$.
Therefore, only ``real'' spin dependence of the strong interaction (e.g. due
to $\pi$-meson exchange) contributes to the IPNCI.

\section{Comparison of the IPNCI with the residual two-body weak
interaction and discussion}
\label{sec:level3}

It is interesting to compare the IPNCI with the initial
two-nucleon weak interaction
$:\hat W(1,2):$.
The interaction (18,19) and the ``bare'' one, Eq.(10,8), have different
isotopic and coordinate structure
(momentum ${\bf p}$ or derivative $\nabla$ instead of radius-vector ${\bf r}$).
Taking into account that $r \sim r_{0}A^{1/3}, p_{F}r \sim p_{F}r_{0}A^{1/3}
\sim A^{1/3}$, we obtain
\begin{equation}
\frac{V^{IPNCI}}{:W(1,2):} \sim
p_{F}r \sim A^{1/3}.
\end{equation}
For heavy nuclei where neutron-nucleus PNC effects were
measured the nucleon number $A \simeq 114...240$, and
$r_{0} =1.15 fm \sim p_{F}^{-1}$ is internucleon distance.
Thus, the IPNCI (18,19) is an order of
magnitude stronger than the initial weak interaction (10) acting within
the valence shell.
The numerical results for the matrix elements of $V_{IPNCI}$
as compared to those of the initial interaction $:W:$ between valence
shell states for Th-U region are presented in the Table (${\tilde V}_{IPNCI}$
takes into account the momentum dependent component of the Landau-Migdal
interaction, see below).
In practical calculations, it is useful to treat $V_{IPNCI}$ in the
second quantization form
using multipole expansion in the particle-hole channel:
$V_{IPNCI}=
\frac{1}{2}\sum_{J}((a^{+}b)_{J}V^{IPNCI,J}_{abcd}(c^{+}d)_{J})_{0}$
where
$(...)_{J}$ means the coupling of nucleon creators $a^{\dagger}$ and
destructors $a$ to a given angular momentum $J$ \cite{BM}.
The values of the parameters $g^{W}_{ik}$ and
$g^{W}_{p}$,$g^{W}_{n}$ were chosen according to
\cite{pst},\cite{we}. On the average, the enancement (14)
is an order of magnitude.

We stress once more that the selection rules (change of parity
and conservation of the angular momentum) forbid matrix elements
of the single-particle weak potential between the valence
orbitals presented in the Table, i.e. the IPNCI and the residual
interaction $:W:$ are the only source of parity nonconservation
in the compound states within the ``principal component'' approach.
The equations expressing the root mean square matrix element between
compound states in terms of matrix elements of the IPNCI
and $:W:$ (see the Table) are presented in Ref. \cite{we}.

Of course, the explicit form of IPNCI (Eqs.(18),(19)) based on the
approximation (5) is semiquantitative.
In particular, due to the smallness of the quantity
$h-h'$, corrections to (18) may be considerable for particular matrix
elements.
Especially big corrections appear in the interference term
(proportional to $g^{W}_{p}g^{W}_{n}$) in the calculation of the mean
squared value of the weak matrix element between compound states. These
matrix elements contain a sum of the products of the matrix elements
between the nucleon orbitals (see Eq.(7) and Ref. \cite{we} for
the accurate formula):
\begin{displaymath}
\overline{ |<s|W|p>|^{2}} \sim \sum V^{IPNCI}_{abcd}
V^{IPNCI}_{cdab} \sim
\sum V_{S,ibcd} V_{S,ajcd} w_{ai} w_{jb} + ...
\end{displaymath}
The coefficients before $(g^{W}_{p})^{2}$ and $(g^{W}_{n})^{2}$
in this sum are positive, and the result is stable.
On the other hand, the coefficients before the interference
term ($\sim g^{W}_{p} g^{W}_{n}$) are not positively defined
and this coefficient tends to decrease
after the summations (in comparison with the ones before
$(g^{W}_{p})^{2}$ and $(g^{W}_{n})^{2}$).
Therefore, the result for the mean squared
matrix element is proportional to $|g^{W}_{p}|^{2}+|g^{W}_{n}|^{2}$
with a small coefficient before $g^{W}_{p}g^{W}_{n}$ rather than
to $(g^{W}_{p}-g^{W}_{n})^{2}$ (as it could follow from the
approximate formula (18) for the IPNCI).
%
%

The numerical calculation of the root mean square matrix elements between
compound states has shown that the contribution of the IPNCI (Eqs.(18-19)) is
about $7...12$
times bigger
than the contribution of the initial weak interaction $W$ (Eq.(10))
confirming the estimate (20).

As it was mentioned above
the results (13)-(19) can be obtained using perturbation theory
considerations (see Eqs.(1)-(7)).
Formally, the result in Eq.(6) is obtained in the first order
in residual strong interaction
$V_{S}$. However, iterations ($w_{1} \rightarrow V_{IPNCI}
\rightarrow w_{1}+ \delta w_{1} \rightarrow V_{IPNCI} + \delta V_{IPNCI}
\rightarrow ...$)
of the contribution of the velocity-independent part of the
interaction $V_{S}$ does
change the result since $V_{IPNCI}$ does not contribute to the weak
potential ($<V_{IPNCI}>_{core}=<[A,V_{S}]>_{core}=0$, see Eq.(7)).
This explains why ``all-orders'' results (13)-(19)
coincide with the first order result (6): the self-consistent
random-phase-approximation-like
chain is terminated after the first iteration.
The situation changes if one takes into
account the momentum-dependent corrections to the Landau-Migdal interaction
given by \cite{Brown},\cite{SAPERSHTEIN}. In this case, the
summation of the series
produces an additional enhancement factor $\sim 1.5$.

\section{Contribution of the velocity dependent residual strong
interaction to the renormalization of the weak potential and the IPNCI}
\label{sec:level4}

Let us consider now these momentum dependent corrections $V_{1}$
to the Landau-Migdal interaction (16),
given by
\begin{eqnarray}
V_{1}= \frac{1}{4} C p_{F}^{-2} (h_{1}+h'_{1} {\bf \tau}_{1}
{\bf \tau}_{2})
({\bf \sigma}_{1}{\bf \sigma}_{2}) \times
\nonumber\\
\times
[ {\bf p}_{1} {\bf p}_{2} \delta({\bf r}_{1}-{\bf r}_{2}) +
{\bf p}_{1} \delta({\bf r}_{1}-{\bf r}_{2}) {\bf p}_{2} +
{\bf p}_{2} \delta({\bf r}_{1}-{\bf r}_{2}) {\bf p}_{1} +
\delta({\bf r}_{1}-{\bf r}_{2}){\bf p}_{1} {\bf p}_{2} ].
\end{eqnarray}
This form originates from the $\pi$-meson exchange contribution
to the nucleon-nucleon interaction \cite{Brown},\cite{SAPERSHTEIN}.
Its constants are known
to be $h_{1}=-0.5$, $h'_{1}=-0.26$ (Ref. \cite{SAPERSHTEIN}).
Note, that we keep in (21) only those ${\bf p}$-dependent
corrections which yield nonzero contributions to the P-odd
field renormalization (see below). To the lowest powers of ${\bf p}$,
these terms should be
$\sim {\bf \sigma}_{1} {\bf p}_{1} {\bf \sigma}_{2} {\bf p}_{2}$.
Spin-independent velocity contributions to (21) responsible,
e.g., for the effective mass renormalization, are therefore
irrelevant and the effects caused by them (e.g., effective mass
renormalization) are assumed to be taken into account in
definition of the
constants $C$, $m$ and $h_{i}$.
%

It is easy to see, that in this case the operator ${\hat A}$
should be of the same
form as ${\hat a}$, but with its constants renormalized.
The inclusion of the additional term $V_{1}$ (21) ($V_{S}=V+V_{1}$)
gives the ``compensation
equation'' (14) for the effective single-particle field in the form
\begin{displaymath}
\hat w(1) +
\left[
i \sum_{a=1,2} \xi_{a} {\bf \sigma}_{a} {\bf r}_{a},
\frac{{\bf p}^{2}}{2m}
\right]
+ K \{ ({\bf \sigma p}),\rho \}=0,
\end{displaymath}
where
$K= - \frac{C}{2p_{F}^{2}} \lbrack \frac{Z}{A} (h_{1}
\pm h'_{1})\xi_{p}+
\frac{N}{A} (h_{1} \mp h'_{1})\xi_{n} \rbrack$ with upper (lower) signs
for proton (neutron)
respectively (see Ref. \cite{PRC}).
In the constant density approximation,
all terms in this equation have the same operator structure
and its solution is equivalent to the renormalization of the
constants $\xi$ in (11),
obtained by replacement of ``bare'' weak constant $g^{W}_{p,n}$ by their
renormalized values $\tilde g^{W}_{p,n}$:
\begin{eqnarray}
\tilde g^{W}_{p}=\frac{1}{D} \lbrace
g^{W}_{p} [1+ \frac{2N}{3A}(h_{1}+h'_{1})] -
\frac{2N}{3A}g^{W}_{n}(h_{1}-h'_{1}) \rbrace , \nonumber\\
\tilde g^{W}_{n}=\frac{1}{D} \lbrace
g^{W}_{n} [1+ \frac{2Z}{3A}(h_{1}+h'_{1})] -
\frac{2Z}{3A}g^{W}_{p}(h_{1}-h'_{1}) \rbrace,
\end{eqnarray}
with
$D=[1+\frac{2N}{3A}(h_{1}+h'_{1})][1+\frac{2Z}{3A}(h_{1}+h'_{1})]-
\frac{4NZ}{9A^{2}}(h_{1}-h'_{1})^{2}$
(firstly, this result has been obtained in our work \cite{PRC}).
Thus, with the account for
$V_{1}$, the IPNCI
takes the form
\begin{eqnarray}
V_{IPNCI}= {\tilde V}_{IPNCI}+V^{vel}_{IPNCI}= \qquad \qquad \nonumber\\
=2(\tilde \xi_{n} - \tilde \xi_{p})(h'-h)C({\bf \sigma}_{p} [{\bf \sigma}_{n}
\times {\bf r}]) \delta({\bf r}_{p}-{\bf r}_{n})
+ V^{vel}_{IPNCI},
\end{eqnarray}
where the first term has the form of (18) but with
the renormalized constants
$\tilde \xi_{p}$, $\tilde \xi_{n}$, which yields an additional enhancement
at the negative values of $h_{1}$, $h'_{1}$
($\tilde \xi \sim 1.5 \xi$ for $h_{1}=-0.5$, $h'_{1}=-0.26$, see the Table).
Note that at present there is an uncertainty in the
values of $h_{1}$, $h'_{1}$. In Ref. \cite{PRC}, we carried out
one more calculation of the weak potential renormalization basing on the
underlying ($\pi + \rho$)-exchange strong interaction which also
produces a tensor contribution to $V_{1}$.
These calculations give even more substantial enhancement of the
weak potential constants $\tilde{g^{W}}$,$\tilde{\xi}$.

The second term contains velocity dependent
corrections:
\begin{eqnarray}
V^{vel}_{IPNCI}= \qquad :[A,V_{1}]: =\qquad \qquad \qquad \qquad \qquad \qquad
\qquad \qquad \qquad
\nonumber\\
-\frac{C}{4p^{2}_{F}}
[
: \left( (\tilde \xi_{n} + \tilde \xi_{p})(h_{1}+h'_{1}\tau_{1}\tau_{2})
+ \frac{1}{2}(\tilde \xi_{n} - \tilde \xi_{p})(h_{1}+h'_{1})
(\tau_{1z}+\tau_{2z}) \right) \{ ( {\bf \sigma}_{1} {\bf p}_{1})+
({\bf \sigma}_{2} {\bf p}_{2}), \delta({\bf r}_{1}-{\bf r}_{2}) \}
\nonumber\\
+
\frac{1}{2}(\tilde \xi_{n} - \tilde \xi_{p})(h_{1}-h'_{1})
(\tau_{2z}-\tau_{1z})  \{ ( {\bf \sigma}_{1} {\bf p}_{1})-
({\bf \sigma}_{2} {\bf p}_{2}), \delta({\bf r}_{1}-{\bf r}_{2}) \} :
\nonumber\\
+
(\tilde \xi_{n} - \tilde \xi_{p})(h_{1}-h'_{1})
(\tau_{2z}-\tau_{1z}) \{ {\bf p}_{1}, \{ {\bf p}_{2},
( {\bf \sigma}_{1} [ {\bf \sigma}_{2} {\bf r}])
\delta({\bf r}_{1}-{\bf r}_{2}) \} \} ], \qquad \qquad \qquad
\nonumber
\end{eqnarray}
where $\frac{C}{4p_{F}^{2}} \xi = \frac{1}{3} \frac{Gg}{2 \sqrt{2} m}$
and $V^{vel}_{IPNCI}$ (except the last term) has no enhancement
in comparison with the two-body
weak interaction (10). Thus, it is considerably smaller than the first term in
Eq.(23). The last term in $V^{vel}_{IPNCI}$ is in fact the momentum dependent
correction to IPNCI (eqs.(18),(23)).


\section{Contribution of the IPNCI to the regular PNC effects}
\label{sec:level5}

In principle, the IPNCI can also give some regular PNC-effect
in the neutron capture, besides the main ``random'' one \cite{we}.
Consider
the neutron capture into a compound state of negative parity ($p$-wave
compound
resonance for the positive parity target nucleus).
The strong residual interaction can capture the neutron in the $p$-wave only.
The IPNCI (Eqs.(18,19,23))
can capture the $s$-wave neutron.
The slow neutron wave function $exp(i{\bf k r}) \chi
\simeq (1+i({\bf k r})) \chi$
($\chi$ is the spinor) contains both $s$-wave and $p$-wave parts which
are connected by
the relation $\psi_{p1/2}=\frac{ik}{3}({\bf \sigma r})\psi_{s}$.
It is clear that
the IPNCI contribution to the $s$-wave neutron capture
matrix element proportional to
$Q({\bf r}{\bf \sigma}_{p} \times {\bf \sigma}_{n}) \psi_{s}$ (see Eq.(18))
is similar to the spin-dependent part of the $p_{1/2}$-wave strong
contribution $({\bf \sigma}_{p} {\bf \sigma}_{n}) \psi_{p1/2} \rightarrow
\frac{k}{3}({\bf r}{\bf \sigma}_{p} \times {\bf \sigma}_{n}) \psi_{s}+...$
(i.e. $V_{IPNCI} \psi_{s} \sim V_{S} \psi_{p}$).
The similarity of these two fields means that $s$-wave and $p$-wave
neutrons can
excite the same state of the nucleus, and there is a coherent contribution
to the PNC effects (which is proportional to the doubled ratio of the
$s$-wave
to $p_{1/2}$-wave capture amplitudes):
\begin{displaymath}
P \sim 2\frac{T_{s}}{T_{p1/2}} \sim 6\frac{\tilde \xi}{k}
=\frac{1.3 \cdot 10^{-3} \tilde g^{W}_{n}}{\sqrt{E}}.
\end{displaymath}
Here, $E$ is neutron energy in eV. This value is comparable with the valence
contribution estimates \cite{te1},\cite{te2},\cite{STODOLSKY},
if $\tilde g^{W}_{n} \sim g^{W}_{n} \sim 1$ (small renormalization
of P-odd field)
and it is
too small in comparison to the observed regular effect in neutron
capture by $^{232}Th$
($P \simeq 0.3/\sqrt{E}$).

\section{Induced parity and time invariance violating interaction}
\label{sec:level6}

It is interesting to compare the IPNCI with a similar parity and time
invariance violating interaction (IPTI) which is induced by the strong
interaction
$V_{S}$ and P,T-odd nuclear potential (instead of P-odd potential (3)):
\begin{displaymath}
W_{PT}=
\eta_{PT} \frac{G}{2 \sqrt{2} m} ({\bf \sigma \nabla}) \rho \simeq
- \lambda ({\bf \sigma \nabla}) U,
\end{displaymath}
with $\lambda = \eta_{PT} \frac{G}{2 \sqrt{2} m} \frac{ \rho (0)}{|U(0)|} =
2 \cdot 10^{-8} \eta \cdot fm$.
The shape of the strong potential $U$ and that of the
nuclear density are assumed
to be similar. The wave function perturbed by this interaction
(see Ref. \cite{ZHETF}) can be written as
${\tilde \psi} = exp(-{\hat A}_{PT}) \psi \simeq (1-{\hat A}_{PT}) \psi$,
${\hat A}_{PT}=\lambda ({\bf \sigma \nabla})$. Calculations similar to
those we have done for the IPNCI give the following result
\begin{displaymath}
V^{IPTI}=[{\hat A}_{PT},V_{s}] \sim
C \lambda [({\bf \sigma \nabla}),\delta ({\bf r}_{1}-{\bf r}_{2})]
-
i C \lambda ( {\bf \sigma}_{1} \times {\bf \sigma}_{2} \{ {\bf \nabla}_{1},
\delta ({\bf r}_{1}-{\bf r}_{2}) \} )
\end{displaymath}
with $C \lambda = C \frac{\rho}{|U|} \frac{G}{2 \sqrt{2} m} \eta_{PT} \simeq
\frac{G}{2 \sqrt{2} m} \eta$. We see that the structure
and strength of IPTI is similar to those of the initial
two-body P,T-odd interaction
(see e.g., \cite{nov3}).
Thus, in the case of P,T-odd interaction there is no $A^{1/3}$ enhancement.

\section{Conclusion}
\label{sec:level7}

Let us stress in conclusion, that we considered here the IPNCI term in
the Hamiltonian
caused by the change of the residual
strong interaction by the coherent PNC field. An explicit expression for the
IPNCI is obtained. It is shown that the IPNCI
is $\sim A^{1/3}$ times stronger than
residual two-nucleon weak interaction. This enhancement is due to
coherent contributions of all nucleons to the weak
nucleon-nucleus
potential. (In the initial weak interaction, only the two external nucleons
interact, while
in the IPNCI, contributions of all the nucleons into the weak potential are
accumulated).
Of course, the P-even strong interaction $V_{S}$ remains unchanged
in the total Hamiltonian (see Eq.(15)), and its main effect, i.e., mixing
of configurations in true eigenstates, remains to be separate problem.

\section{Acknowledgement}
\label{sec:nolevel}

Useful discussions with I.B.Khriplovich and O.P.Sushkov are acknowledged.
We are grateful to V.F.Dmitriev and V.B.Telitsin
kindly providing us with the code for the nuclear wave functions
calculation.
The work was supported by the Australian Research Council.

\newpage
\widetext
\begin{table}
\caption{Absolute values of the matrix elements of $V_{IPNCI}$ (Eq.(12,13)),
${\tilde V}_{IPNCI}$ (Eq.(18),(14)
with matrix elements $w$ renormalized according to
(17))
and $:W:$ (Eq.(4,2))  in $eV$ for the $Th$-$U$ region.
$a,b$ ($c,d$) denote the single particle neutron (proton)
upper states.}

\begin{tabular}{llcdcc}
$a$ \qquad \qquad $b$ \qquad \qquad $c$ \qquad \qquad  $d$ \qquad J \qquad
$|V^{IPNCI,J}_{abcd}|$ \qquad
$|{\tilde V}^{IPNCI,J}_{abcd}|$ \qquad $|:W:^{J}_{abcd}|$ \\
\tableline
2$g$ 9/2 \quad 1$j$15/2 \quad 1$h$ 9/2 \quad 1$h$ 9/2 \qquad 3
\qquad 0.067 \qquad \qquad \quad 0.082 \qquad \quad \qquad 0.009 \\
2$g$ 9/2 \quad   1$j$15/2  \quad  1$h$ 9/2 \quad   1$h$ 9/2 \qquad  4
 \qquad 0.033 \qquad \quad \qquad 0.062 \qquad \quad \qquad 0.001 \\
2$g$ 9/2 \quad   1$j$15/2 \quad   1$h$ 9/2 \quad   1$h$ 9/2 \qquad  5
 \qquad 0.035 \qquad \quad \qquad 0.048 \qquad \quad \qquad  0.012 \\
2$g$ 9/2 \quad   1$j$15/2 \quad   1$h$ 9/2 \quad   1$h$ 9/2 \qquad  7
 \qquad 0.029 \qquad \quad \qquad 0.043 \qquad \quad \qquad  0.016  \\
2$g$ 9/2 \quad   1$j$15/2 \quad   1$h$ 9/2 \quad   1$h$ 9/2 \qquad  8
 \qquad 0.043 \qquad \quad \qquad 0.082 \qquad \quad \qquad 0.001 \\
1$i$11/2 \quad   1$j$15/2 \quad   1$h$ 9/2 \quad   1$h$ 9/2 \qquad  3
 \qquad 0.144 \qquad \quad \qquad 0.184 \qquad \quad \qquad 0.007 \\
1$i$11/2 \quad   1$j$15/2 \quad   1$h$ 9/2 \quad   1$h$ 9/2 \qquad  5
 \qquad 0.130 \qquad \quad \qquad 0.165 \qquad \quad \qquad 0.016 \\
1$i$11/2 \quad   1$j$15/2 \quad   1$h$ 9/2 \quad   1$h$ 9/2 \qquad  7
 \qquad 0.131 \qquad \quad \qquad 0.166 \qquad \quad \qquad 0.032 \\
1$i$11/2 \quad   1$j$15/2 \quad   1$h$ 9/2 \quad   1$h$ 9/2 \qquad  9
\qquad 0.172  \qquad \quad \qquad 0.218  \qquad \quad \qquad 0.027
\end{tabular}
\end{table}
\noindent

\begin{thebibliography}{200}
\bibitem{te0} J.D.~Bowman, C.D.Bowman, J.E.Bush, P.P.J.Delheij,
C.M.Frankle, C.R.Gould, D.G.Haase, J.Knudson, G.E.Mitchel, S.Penttila,
H.Postma, N.R.Robertson, S.J.Seestrom, J.J.Szymansky, V.W.Yuan, and X.Zhu,
Phys. Rev. Lett. {\bf 65}, 1192 (1990),
V.W.~Yuan {\it et al.}, Phys.Rev. {\bf C44}, 2187 (1991),
C.M.~Frankle, J.D.~Bowman, J.E.Bush, P.P.J.Delheij,C.M.Frankle, C.R.Gould,
D.G.Haase, J.Knudson, G.E.Mitchel, S.Penttila, H.Postma, N.R.Robertson,
S.J.Seestrom, J.J.Szymansky, S.H.Yoo, V.W.Yuan,
and X.Zhu,
Phys. Rev. Lett. {\bf 67}, 564 (1991).
\bibitem{te1}
M.B.Johnson, J.D.Bowman, and S.H.Yoo, Phys.Rev.Lett. {\bf 67}, 310 (1991).
\bibitem{te11}
J.B.French, V.K.B.Kota, A.Pandey and S.Tomsovic,
Ann.Phys. (N.Y.) {\bf 181}, 235(1988);
{\bf 181}, 198(1988); H.A.~Weidenm\"uller, Nucl.Phys. {\bf A522}, 293c (1991);
V.V.Flambaum, Phys.Rev. {\bf C45}, 437 (1992).
\bibitem{te2} A.M\"uller, E.D.Davis, and H.L.Harney, Phys.Rev.Lett. {\bf 65},
1329 (1990),
J.D.~Bowman, G.T.Garvey, C.R.Gould, A.C.Hayes, and M.B.Johnson,
Phys.Rev.Lett., {\bf 68}, 780 (1992);
N.Auerbach, Phys.Rev. {\bf C45}, R514 (1992); S.E.Koonin, C.W.Johnson, and
P.Vogel, Phys.Rev.Lett. {\bf 69}, 1163 (1992);
N.Auerbach and J.D.Bowman, Phys.Rev. {\bf C46}, 2582 (1992);
J.D.Bowman {\it et al},
%
in:
{\it Time Reversal Invariance and Prity Violation in Neutron
Reactions}, Wrld.Sci., Singapore, 1993, p.8.

\bibitem{LW} C.Lewenkopf and H.A.Weidenm\"uller,
Phys.Rev. {\bf C46}, 2601 (1992).
\bibitem{MASUDA} Y.Masuda, T.Adachi, A.Masaike and K.Morimoto,
Nucl.Phys. {\bf A504}, 269 (1989).
\bibitem{A1} V.V.Flambaum and I.B.Khriplovich, Zh.Eksp.Teor.Fiz.
{\bf 79},1656(1980) [Sov.Phys. JETP {\bf 52}, 835(1980)].
\bibitem{A2} W.C.Haxton and E.M.Henley, M.J.Musolf, Phys.Rev.Lett.
{\bf 63}, 949(1989).
\bibitem{A3} C.Bouchiat and C.A.Piketty, Z.Phys. {\bf C49}, 91 (1991).
\bibitem{we} V.F.~Flambaum and O.K.~Vorov, Phys.Rev.Letts. {\bf 70},4051
(1993).

\bibitem{MITCHEL}
J.F.Shriner, Jr., and G.E.Mitchel, Phys.Rev. {\bf C49}, R616 (1994).

\bibitem{BM} A.~Bohr and B.~Mottelson, {\it Nuclear Structure} (Benjamin,
New York, 1969), Vol. 1.
\bibitem{Zar} D.F.~Zaretsky and V.I.~Sirotkin, Yad. Fiz. {\bf 37}, 607 (1983)
[Sov. J. Nucl. Phys. {\bf 37}, 361 (1983)]; {\bf 45}, 1302 (1987) [{\bf 45},
808 (1987)].
\bibitem{kad} S.G.Kadmensky, V.P.~Markushev, and V.I.~Furman, Yad. Fiz.
{\bf 37}, 581 (1983) [Sov. J. Nucl. Phys. {\bf 37}, 345 (1983)].

\bibitem{Nog} S.~Noguera and B.~Desplanques, Nucl. Phys. {\bf A457},
189 (1986); B.~Desplanques and S.~Noguera, Nucl.Phys. {\bf A561},
189 (1993).

\bibitem{DDH} B.~Desplanques, J. Donoghue and B.~Holstein, Ann. of Phys.
{\bf 124},449 (1980).
\bibitem{ADELBERGER} E.G.~Adelberger and W.C.~Haxton, Ann. Rev. Nucl. Part.
Sci. {\bf 35}, 501 (1985).
\bibitem{GELLMANN} R.P.Feynman and M.Gell-Mann, Phys.Rev. {\bf 109},193
(1958).
\bibitem{nov3} V.V.~Flambaum, I.B.~Khriplovich, and O.P.~Sushkov,
Nucl. Phys. {\bf A449}, 750 (1986).
\bibitem{MCKELLAR}
B.H.J.McKellar, Phys.Rev.Lett. {\bf 21}, 1822 (1968).

\bibitem{nov2} V.V.~Flambaum, I.B.~Khriplovich, and O.P.~Sushkov,
Phys.Lett. {\bf 146B}, 367 (1984).

\bibitem{nov22} V.F.Dmitriev, V.V.Flambaum, O.P.Sushkov,
and V.B.Telitsin, Phys.Lett., {\bf 125B}, 1 (1983).

\bibitem{STODOLSKY} B.Desplanques,
Talk at the International Symposium on Weak and Electromagnetic
interactions in Nuclei, Dubna, September 1990;
L.Stodolsky, Phys.Lett. {\bf 50B}, 352 (1974);
Phys.Lett. {\bf 96B}, 127 (1980),
G.Karl and D.Tadic, Phys.Rev. {\bf C16}, 1726 (1977).

\bibitem{PRC} V.V.Flambaum and O.K.Vorov, Phys.Rev. {\bf C49}, 1827 (1994).
\bibitem{Brown} G.E.Brown, Rev.Mod.Phys., {\bf 43},1 1971);
V.Klemt, S.A.Moszkowski, and J.Speth, Phys.Rev. {\bf C14}, 302 (1976);
J.Speth, E.Werner, and W.Wild, Phys.Rep. {\bf 33}, No.3, 127(1977) and
references therein;
G.Bertsch, D.Cha, and H.Toki, Phys.Rev. {\bf C24}, 533 (1981);
J.W.Negele, Rev.Mod.Phys. {\bf 54}, 913 (1982);
R.De Haro, S.Krewald, and Speth, Nucl.Phys., {\bf A388},
265 (1982); K. Goeke, and J.Speth, Annu. Rev. Nucl. Part. Sci. {\bf 32}, 65
(1982); F.~Osterfeld, Rev.Mod.Phys., {\bf 64}, 491 (1992), and references
therein;
A.Green, W.Unkelbach, F.T.Baker, D.Beatty, L.Bimbot, X.Y.Chen,
V.R.Cupps, C.Djalali, G.Edwards, R.W.Fergerson, C.Glashausser,
K.W.Jones, M.K.Jones, G.Kumbartzki, A.Sethi, B.Storm
and J.Wambach,
Phys.Rev.Lett. {\bf 70}, 734(1993).
\bibitem{SAPERSHTEIN}
V.A.Khodel and E.E.Sapershtein, Phys.Rep. {\bf 92}, 183 (1982), and
references therein;
H.Nopre and E.Werner, Z.Phys. {\bf 254}, 345 (1972).
\bibitem{BACKMAN} S.-O. B\"akman, O.Sj\"oberg and A.D.Jackson,
Nucl.Phys. {\bf A321}, 10 (1979);
G.E.Brown, S.-O. B\"akman, E.Oset and W.Weise,
Nucl.Phys. {\bf A286}, 191 (1977);
J.Speth, V.Klemt, J.Wambach and G.E.Brown,
Nucl.Phys. {\bf A343}, 382 (1980).
\bibitem{Migdal} A.B.Migdal, {\it Theory of Finite Fermi Systems and
Applications to atomic Nuclei} (John Wiley \& Sons, New York, 1967).
\bibitem{nov1} F.C.~Michel, Phys.Rev. {\bf B133}, 329 (1964).

\bibitem{Landau} L.D.Landau, Zh.Eksp.Theor.Fiz. {\bf 30}, 1058 (1956)
[Sov.Phys.JETP {\bf 30}, 920 (1956)]; Zh.Eksp.Theor.Fiz. {\bf 32}, 59 (1957)
[Sov.Phys.JETP {\bf 5}, 101 (1957)]; Zh.Eksp.Theor.Fiz. {\bf 35}, 97 (1958)
[Sov.Phys.JETP {\bf 8}, 70 (1959)].
\bibitem{OKUN} L.B.Okun, {\it Leptons and Quarks} (North Holland Publishing
Co., Amsterdam 1982).
\bibitem{pst} A.P.Platonov, Yad.Phys. {\bf 42}, 361 (1984) [Sov. J. Nucl.
Phys. (1984)];
O.P.~Sushkov and V.B.~Telitsin, Phys.Rev. {\bf C48}, 1069 (1994).
\bibitem{ZHETF} O.P.~Sushkov, V.V.~Flambaum and I.B.Khriplovich,
Zh.Eksp.Theor.Fiz. {\bf 87},
521 (1984) [Sov.Phys.JETP {\bf 60}, 873 (1984)].
%





\end{thebibliography}
\end{document}